\documentclass[reprint,
 amsmath,amssymb,
 aps,
floatfix,
]{revtex4-1}

\usepackage{graphicx}
\usepackage{dcolumn}
\usepackage{bm}

\usepackage[
            pdfstartview=FitH,
            bookmarksnumbered=true,
            bookmarksopen=true,
            colorlinks,
            linkcolor=yellow,
            anchorcolor=green,
            citecolor=red
            ]{hyperref}
\usepackage[toc,page,title,titletoc,header]{appendix}
\newcommand{\be}{\begin{equation}}
\newcommand{\ee}{\end{equation}}
\newcommand{\bea}{\begin{eqnarray}}
\newcommand{\eea}{\end{eqnarray}}

\usepackage{color}

\def\ra{\rangle}

\def\bea{\begin{eqnarray}}
\def\eea{\end{eqnarray}}
\def\be{\begin{equation}}
\def\ee{\end{equation}}

\def\Tr{\mathrm{Tr}}

\newcommand{\ket}[1]{\left| {#1} \right\rangle}
\newcommand{\bra}[1]{\left\langle {#1} \right|}

\newcommand{\bikete}[2]{\left| {#1} \right\rangle_{\text{R}}\left| {#2} \right\rangle_{\bar {\text{R}}}}

\newcommand{\pa}{p}

\newcommand{\rom}[1]{\textup{\uppercase\expandafter{\romannumeral#1}}}

\usepackage[utf8]{inputenc}
\usepackage[english]{babel}
\usepackage{amsthm}

\def\slashchar#1{\setbox0=\hbox{$#1$} 
\dimen0=\wd0 
\setbox1=\hbox{/} \dimen1=\wd1 
\ifdim\dimen0>\dimen1 
\rlap{\hbox to \dimen0{\hfil/\hfil}} 
#1 
\else 
\rlap{\hbox to \dimen1{\hfil$#1$\hfil}} 
/ 
\fi}

\begin{document}


\title{Holevo Bound of Entropic Uncertainty in Schwarzschild Spacetime}

\author{Jin-Long Huang}
\thanks{These three authors contributed equally}
\affiliation{Department of Physics, Southern University of Science and Technology , Shenzhen, 518055, China}

\author{Wen-Cong Gan}
\thanks{These three authors contributed equally}
\affiliation{
Department of Physics, Nanchang University, No. 999 Xue Fu Avenue, Nanchang, 330031, China}
\affiliation{
Center for Relativistic Astrophysics and High Energy Physics, Nanchang University, No. 999 Xue Fu Avenue, Nanchang 330031, China}

\author{Yunlong Xiao}
\thanks{These three authors contributed equally}
\affiliation{Department of Mathematics and Statistics and Institute for Quantum Science and Technology,
University of Calgary, Calgary, Alberta T2N 1N4, Canada}

\author{Fu-Wen Shu}
\thanks{Corresponding author}
\thanks{shufuwen@ncu.edu.cn}
\affiliation{
Department of Physics, Nanchang University, No. 999 Xue Fu Avenue, Nanchang, 330031, China}
\affiliation{
Center for Relativistic Astrophysics and High Energy Physics, Nanchang University, No. 999 Xue Fu Avenue, Nanchang 330031, China}

\author{Man-Hong Yung}
\thanks{Corresponding author: yung@sustc.edu.cn}
\affiliation{Institute for Quantum Science and Engineering and Department of Physics, Southern University of Science and Technology , Shenzhen, 518055, China}
\affiliation{
Shenzhen Key Laboratory of Quantum Science and Engineering, Shenzhen 518055, China}

\date{\today}

\begin{abstract}
For a pair of incompatible quantum measurements, the total uncertainty can be bounded by a state-independent constant. However, such a bound can be violated if the quantum system is entangled with another quantum system (called memory); the quantum correlation between the systems can reduce the measurement uncertainty. On the other hand, in a curved spacetime, the presence of the Hawking radiation can increase the uncertainty in quantum measurement. The interplay of quantum correlation in the curved spacetime has become an interesting arena for studying quantum uncertainty relations.  Here we demonstrate that the bounds of the entropic uncertainty relations, in the presence of memory, can be formulated in terms of the Holevo quantity, which limits how much information can be encoded in a quantum system. Specifically, we considered two examples with Dirac fields, near the event horizon of a Schwarzschild black hole, the Holevo bound provides a better bound than the previous bound based on the mutual information. Furthermore, if the memory moves away from the black hole, the difference between the total uncertainty and the Holevo bound remains a constant, not depending on any property of the black hole. 
\end{abstract}

\pacs{Valid PACS appear here}

\maketitle


\section{\label{sec:intro}Introduction}

Traditionally, uncertainty principle in quantum mechanics has been formulated in terms of variance 
\cite{Heisenberg1927, Robertson1929, Schrodinger1930, Abbott2016, Schwonnek2017, Xiao2017I}, while in the context of both 
classical and quantum information sciences, it is more natural to use entropy to quantify uncertainties.
The first entropic uncertainty relation (EUR) for position and momentum was given in \cite{Bialynicki1975}, which can 
be shown to be equivalent to Heisenberg's original relation. Later Deutsch \cite{Deutsch1983} formulated entropic 
uncertainty relation for any pair of observables with bounded spectrums. An improvement of Deutsch's entropic 
uncertainty relation was subsequently conjectured by Kraus \cite{Kraus1987} and later proved by Maassen and 
Uffink \cite{Maassen1988}. However, if the measured system $\rho_A$ is prepared with a quantum memory $\rho_B$, correlation between $\rho_A$ and $\rho_B$ will decrease the entropic uncertainties of $\rho_A$.
The entropic uncertainty relation in the presence of quantum memory was proposed by Berta {\it et al.}: 
\begin{equation}
H(M_{1}|B)+H(M_{2}|B)\geqslant U_1 \ ,
\end{equation}
where $U_1\equiv -\log c_{1}+H(A)-\mathcal{I}(A: B)$ \cite{Berta2010}, where $H(\rho)\equiv -\text{tr}(\rho \text{ ln} \rho)$ is von Neumann entropy for density matrix $\rho$ and $\mathcal{I}(A: B)=H(A)+H(B)-H(A, B)$ is the quantum mutual information.

However, is this mutual information
$\mathcal{I}(A: B)$ capable of depicting \textit{how} the uncertainty would behave when the correlation changes? Using Holevo quantity (or Holevo bound), we generalized entropic uncertainty with quantum memory. Let $\mathcal{J}(B|M_1) \equiv H(B)-\sum_j p^1_j H(\rho_{B|u^1_{j}})$ be the Holevo quantity for Bob about Alice's $M_1$ measurement outcomes. The new uncertainty relation says that 
\begin{equation}\label{uncertainty}
H(M_{1}|B)+H(M_{2}|B)\geqslant U_2 \ ,
\end{equation}
where $U_2\equiv -\log c_{1}+H(A)-\mathcal{J}(B|M_1)-\mathcal{J}(B|M_2)$. The new entropic uncertainty relation has the property that the difference (which we denote as $\Delta_2$ below) between entropic uncertainty (LHS of \eqref{uncertainty}) and the new uncertainty lower bound $U_2$ only depends on incompatible measurements $M_1, M_2$ and $\rho_A$, and independent of quantum memory. 
\begin{equation}
\Delta_2^{M_1 M_2}=H(M_1)+H(M_2)+\log c_1-H(A) \ ,
\end{equation}
This also means that the new lower bound $U_2$ can capture the characteristic how the entropic uncertainty would behave corresponding to different quantum memory $\rho_B$, since the difference between $U_2$ and LHS is always a constant. While for $U_1$ the difference between LHS and $U_1$ may increase or decrease as the quantum memory $\rho_B$ changes. Additionally, the terms $-\mathcal{J}(B|M_1)-\mathcal{J}(B|M_1)$ in the RHS of \eqref{uncertainty} can be further lower bounded according to enhanced Information Exclusion Relations\cite{XiaoSci2016}.

Different examples with Dirac fields states in Schwarzschild spacetime are calculated to demonstrate the Holevo quantity $\mathcal{J}(B|M_1)+\mathcal{J}(B|M_2)$, as a new measure of correlation, is the underlying quantity between the quantum memory and entropic uncertainty, rather than previous measure $\mathcal{I}(A: B)$. In this case quantum states are superposition of vacuum state and excited states of Dirac fields. We also call a quantum state as a mode of Dirac fields. The Hilbert space is four-dimensional with basis $\ket{0},\ket{\uparrow},\ket{\downarrow} \text{and} \ket{p}=\ket{\uparrow \downarrow}$, not two in \cite{Feng:2015xza}: $\ket{0}$ and $\ket{1}$. In a spacetime with a Schwarzschild black hole, we consider the case in which the quantum memory $\rho_B$ hovers near the event horizon outside the black hole and the measured system $\rho_A$ is free falling. It has been known that the entanglement between $\rho_A$ and $\rho_B$ would degrade when $\rho_B$ get closer to the event horizon due to Unruh effect on $\rho_B$\cite{Alice:2005}, so the lower bound for entropic uncertainty would increase\cite{Feng:2015xza}. Because $\Delta_2$ is independent of $\rho_B$, $U_2$ is always a constant away from LHS. In other words, when quantum memory gets closer to the event horizon, the correlation between it and measured system is decreased, and the decreased correlation is equal to increased entropic uncertainty.

\begin{figure}[h]
\centering 
\includegraphics[width=.22\textwidth]{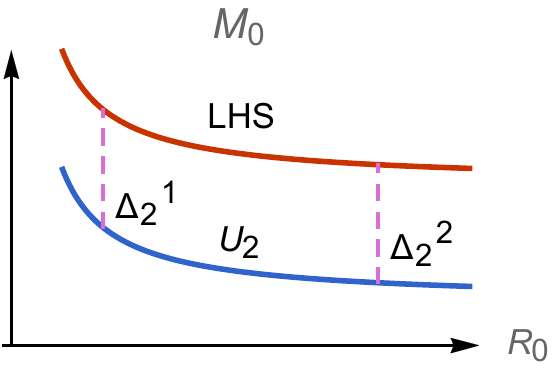}
\hfill
\includegraphics[width=.22\textwidth]{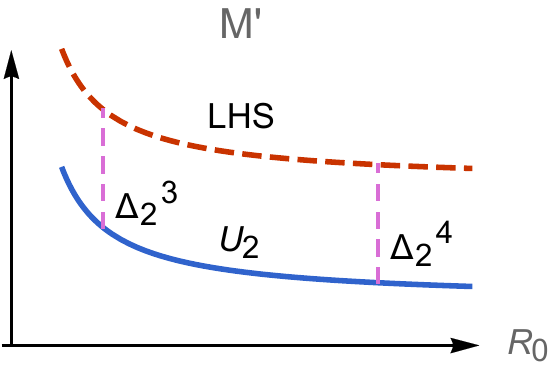}
\caption{Entropic uncertainty (LHS) and lower bound $U_2$ using accessible information for black hole with mass $M_0$ and $M'$. $\Delta_2^1=\Delta_2^2=\Delta_2^3=\Delta_2^4$ where $\Delta_2^1$ and $\Delta_2^2$ correspond to different relative distance $R_0=r_0/2M$ from event horizon. It is sufficient to conduct experiments near \textit{one} Schwarzschild black hole with mass $M_0$ to obtain LHS. For any other Schwarzschild black holes with mass $M'$, we can precisely predict its LHS using LHS for $M_0$, $\Delta_2$ and $U_2$ for $M'$.}
\label{illustration}
\end{figure}

From an experiment point of view, a proper uncertainty game \cite{Prevedel:2011,Li:2011} can be conducted to measure Bob's uncertainty LHS about Alice's measurement outcomes for a particular Schwarzschild black hole with mass $M_0$. $U_2$ and $\Delta_2$ can be calculated for this black hole from Rindler decomposition. Since the difference $\Delta_2$ between LHS and $U_2$ is independent of mass of black hole $M$, energy $\omega$ of quantum state and the relative distance $R_0$ of quantum memory from event horizon, $\Delta_2$ for other black hole can be obtained. Fixing the energy of mode $\omega$, for an arbitrary Schwarzschild black hole with different mass $M'$, we can predict Bob's entropic uncertainty accurately without conducting any other new experiments.

This letter is organized as follows. In Sec.\ref{sec:eur} we propose to use Holevo $\chi$-quantity as a part of lower bound for entropic uncertainty relation with quantum memory, and prove that the difference between two sides of this inequality is independent of quantum memory $B$. Then in Sec.\ref{un-sch} by calculating different examples in Schwarzschild spacetime with quantum states being Dirac field states, we demonstrate that the new lower bound $U_2$ is tighter than previous bound $U_1$, and more importantly, the Holevo quantity $\mathcal{J}(B|M_1)+\mathcal{J}(B|M_2)$ serves as a better correlation measure to reveal how quantum memory would change entropic uncertainty.

\section{Entropic uncertainty relation and Information exclusion principle}\label{sec:eur}
Entropic uncertainty relation proved by Maassen and 
Uffink \cite{Maassen1988} is (we use base $2$ $\log$ throughout this paper),
\begin{align}\label{e:MU}
H(M_{1})+H(M_{2})\geqslant\log \frac1{c_{1}},
\end{align}
where $M_{1}=\{|u_{j}\rangle\}$ and $M_{2}=\{|v_{k}\rangle\}$ are two orthonormal bases on $d$-dimensional Hilbert 
space $\mathcal{H}_{A}$, and $H(M_{1})=-\sum_{j}p_{j}\log p_{j}$ is the Shannon entropy of the probability distribution 
$\{p_{j}=\langle u_{j}|\rho_{A}|u_{j}\rangle\}$ for state $\rho_{A}$ of $\mathcal{H}_{A}$ (similarly for $H(M_{2})$ 
and $\{q_{k}=\langle v_{k}|\rho_{A}|v_{k}\rangle\}$). The number $c_{1}$ is the largest overlap among all 
$c_{jk}=|\langle u_{j}|v_{k}\rangle|^{2}$ ($\leqslant 1$) between $M_{1}$ and $M_{2}$.

EUR \eqref{e:MU} can be improved as \cite{Berta2010}
\begin{align}\label{e:MU2}
H(M_{1})+H(M_{2})\geqslant\log \frac1{c_{1}}+H(A),
\end{align}
where $H(A)$ characterize the amount of uncertainty increased by the mixedness of $A$.

However, if the measured system $A$ is prepared with a quantum memory $B$, then the entropic uncertainties in the presence 
of memory are
\begin{align}
H(M_{1}|B)+H(M_{2}|B),
\end{align}
where $H(M_{1}|B)=H(\rho_{M_1B})-H(\rho_B)$ is the conditional entropy with 
$\rho_{M_{1}B}=\sum_{j}(|u_{j}\rangle\langle u_{j}|\otimes I)(\rho_{AB})(|u_{j}\rangle\langle u_{j}|\otimes I)$ 
(similarly for $H(M_{2}|B)$). Then the difference between $H(M_{1})+H(M_{2})$ and $H(M_{1}|B)+H(M_{2}|B)$, i.e.
\begin{align}
H(M_{1})+H(M_{2})-H(M_{1}|B)-H(M_{2}|B),
\end{align}
reveals the uncertainty decrease due to the correlations between measured system $A$ and quantum memory $B$.

At the heart of information theory lies the mutual information, Shannon's fundamental theorem (\cite{Reza1961}, Chapter 12)
states that the mutual information corresponding to a measurement is the average amount of error-free data which may 
be gained through the measurement of system. Information is a natural tool and concept in communications and physics, 
in an operation of measurement or communication, one may seek to maximize the gained information. This kind of optimization
is trivial for classical systems \cite{Hall1994}.

One route to generalize mutual information is motivated by replacing the classical probability distribution by the density
matrices of quantum systems., e. g.,
\begin{align}
\mathcal{I}(A: B)=H(A)+H(B)-H(A, B).
\end{align}     
Here, $H(A)$ stands for the von Neumann entropy of quantum state $A$
and $H(A, B)$ denotes the information of combined system.

On the other hand, the quantum memory $B$, after the measurement corresponding to $|u_{j}\rangle$ ($M_{1}$) has been preformed,
becomes
\begin{align}
\rho_{B|u_{j}}=\langle u_{j}|\rho_{AB}|u_{j}\rangle/\Tr(\langle u_{j}|\rho_{AB}|u_{j}\rangle),
\end{align} 
with probability $p_{j}=\Tr(\langle u_{j}|\rho_{AB}|u_{j}\rangle)$. $H(\rho_{B|u_{j}})$ is the missing information about quantum
memory. The entropies $H(\rho_{B|u_{j}})$ with weighted probability $p_{j}$ leads to a second quantum generalization of mutual 
information
\begin{align}
\mathcal{J}(B|M_1)=H(B)-\sum\limits_{j}p_{j}H(\rho_{B|u_{j}}).
\end{align}
This quantity reveals the information gained about the quantum memory through the measurement $M_{1}$. The difference between
$\mathcal{I}(A: B)$ and $\mathcal{J}(B|M_1)$ is related to quantum discord \cite{Ollivier2001}.

For any quantum systems, the quantity $H(M_{1})+H(M_{2})-H(M_{1}|B)-H(M_{2}|B)$ describes the uncertainty decrease according to the
extra quantum memory, while on the same time $\mathcal{J}(B|M_1)+\mathcal{J}(B|M_2)$ is the increase of information content
of observables due to quantum memory. What is the relation between the values of uncertainty decrease and information increase in 
the presence of quantum memory? Actually, we can rewrite $\rho_{M_{1}B}=\sum_{j}(|u_{j}\rangle\langle u_{j}|\otimes I)(\rho_{AB})(|u_{j}\rangle\langle u_{j}|\otimes I)$ as $\rho_{M_1 B}=\sum_j p_j \ket{u_j}\bra{u_j} \otimes \rho^j_B$, where $\rho^j_B$ is density matrix for $B$ if Alice measurement outcome is $j$. According to joint entropy theorem \cite{Nielsen2011}, we have $H(\rho_{M_1| B})=H(\rho_{M_1 B})-H(\rho_B)=H(M_1)-\mathcal{J}(B|M_1)$, thus
\begin{align}\label{decrease}
&H(M_{1})+H(M_{2})-H(M_{1}|B)-H(M_{2}|B)\notag\\
=&\mathcal{J}(B|M_1)+\mathcal{J}(B|M_2),
\end{align}
for incompatible observables $M_{1}$ and $M_{2}$. Through this unified equation, we have shown that the increase of information content of quantum observables in the presence
of quantum memory equals to the decrease of quantum uncertainties due to the extra quantum memory. Now these two fundamental concepts
in quantum theory and information theory have been unified.

The entropic uncertainty relation now tells that $H(M_{1}|B)+H(M_{2}|B)\geqslant U_2$, where $U_2\equiv -\log c_{1}+H(A)-\mathcal{J}(B|M_1)-\mathcal{J}(B|M_2)$. The left hand side (LHS) minus right hand side (RHS) equals to $\Delta_2^{M_1 M_2}=H(M_1)+H(M_2)+\log c_1-H(A)$ which is independent of quantum memory $B$. Actually, $\Delta_2$ is the difference between LHS and RHS of \eqref{e:MU2}. Put another way, in the presence of quantum memory, the amount of uncertainty decrease equals to the amount of decrease of the lower bound $U_2$. This fact has been revealed in \eqref{decrease}. In next section, we will apply this Holevo quantity generalized entropic uncertainty relation to the cases with Dirac field states in Schwarzschild spacetime.

\section{Generalized entropic uncertainty relations in Schwarzschild spacetime}\label{un-sch}
\subsection{Setup}
\paragraph{\label{sec:level2}Rindler approximation}
\begin{figure}[h]
\centering 
\includegraphics[width=.45\textwidth]{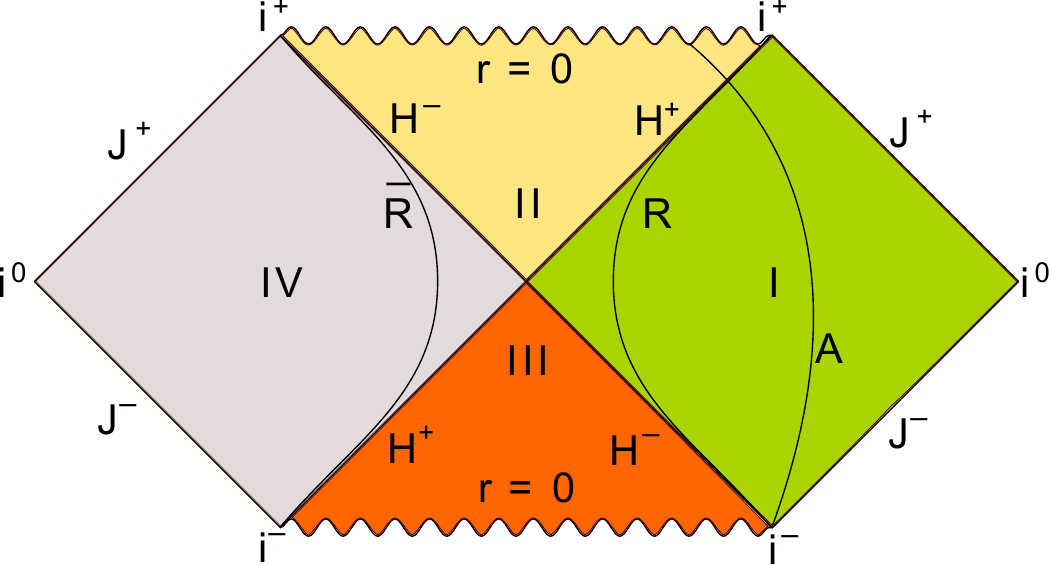}
\caption{Penrose diagram for maximally extended black hole which shows the world-line of Alice, Rob and Anti-Rob. $i^0$ denotes the spatial infinities, $i^-$ ($i^+$) denotes timelike past (future) infinity.  $J^-$ ($J^+$) denotes lightlike past (future) infinity. $H^{\pm}$ denote the event horizons of the black hole. 
}
\label{kruskal}
\end{figure}
\begin{figure}[ht]
\centering 
\includegraphics[width=0.45\textwidth]{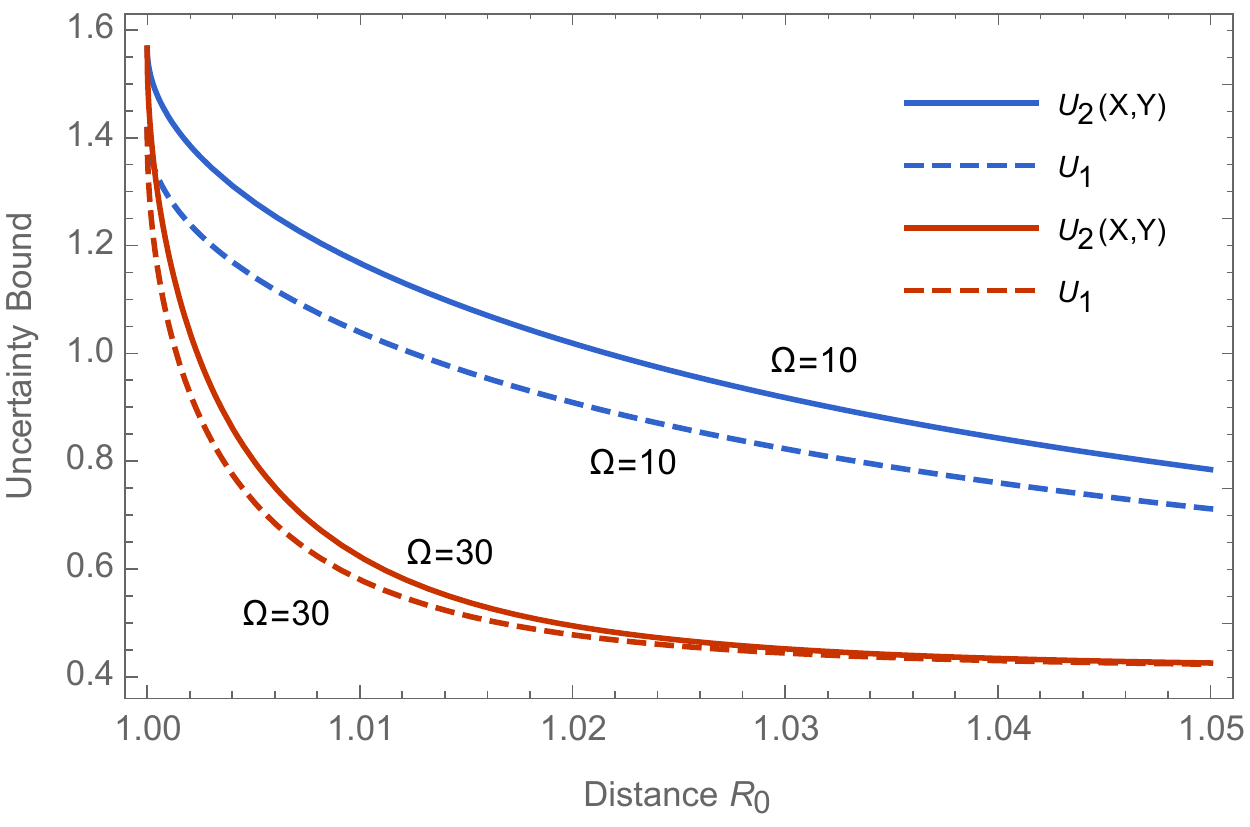}
\hfill
\includegraphics[width=0.45\textwidth]{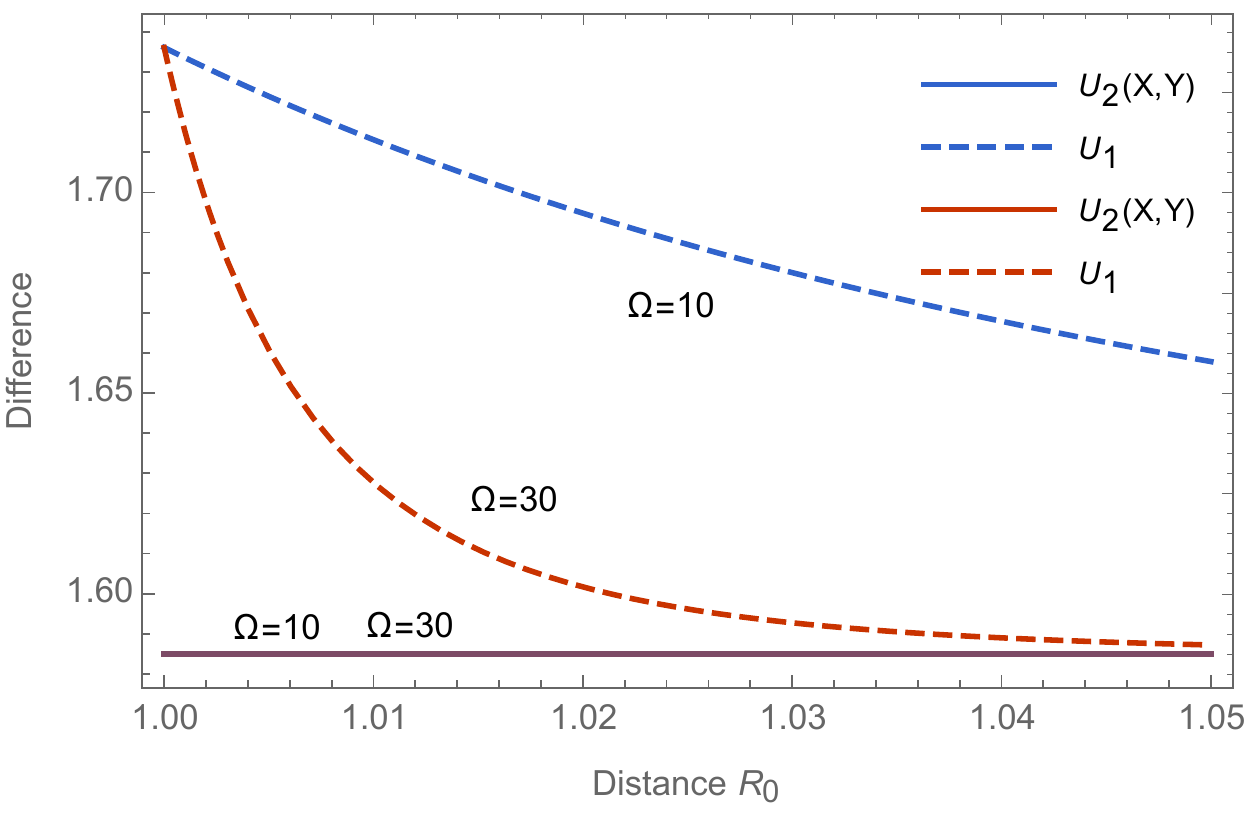}
\caption{We depict EUR lower bound for \eqref{entangleddir} with both $U_1$ and $U_2^{x y}$ and the difference between $H(X|R)+H(Y|R)$. The upper figure depicts the uncertainty bound $U_1$ and $U_2^{x y}$ with respect to $R_0=r_0/2M$ when $\Omega=\omega/T_H=10, 30$. $T_H$ is Hawking temperature and $\omega$ is the frequency of the mode. The lower figure shows the gap between left side and right side: $\Delta_1=H(X|R)+H(Y|R)-U_1$ and $\Delta_2^{x y}=H(X|R)+H(Y|R)-U_2^{x y}$. The relative distance of Rob to event horizon $R_0 \lesssim 1.05$ is assumed thus Rindler approximation can be hold. Given $\Omega=\omega/T_H=10$ or $30$, $U_2^{x y}$ is always better than $U_1$. For different $\Omega$ and $R_0$, $\Delta_2$ is constant while $\Delta_1$ decreases as $\Omega$ and $R_0$ increase. (Colored online.)}
\label{00ud}
\end{figure}
\begin{figure}[ht]
\centering
\includegraphics[width=0.45\textwidth]{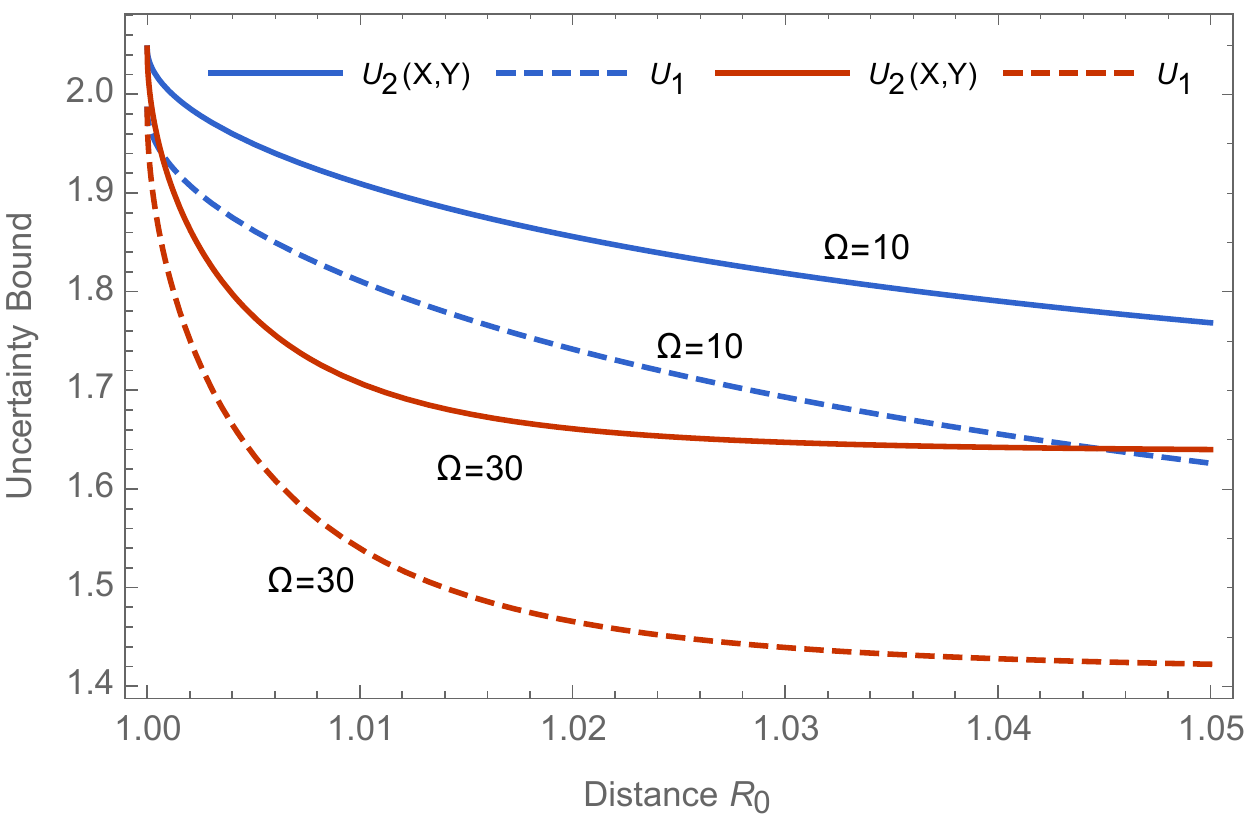}
\hfill
\includegraphics[width=0.45\textwidth]{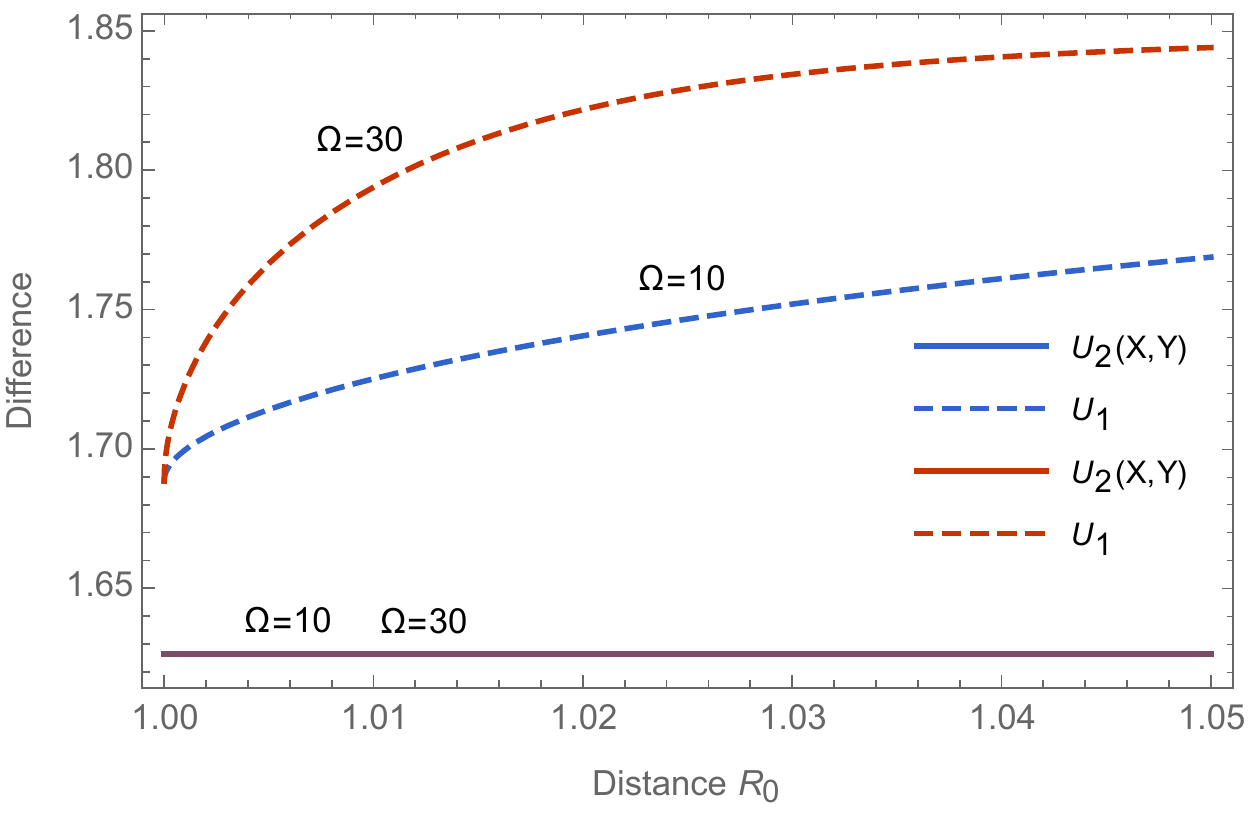}
\caption{Entropy uncertainty bound for Alice and Bob when Alice free falls into the black hole and Bob hovers near the event horizon. When there is no black hole, Alice, Bob and Charlie share a W state $W = \frac{1}{\sqrt{3}} (|00\uparrow \ra+|0\uparrow0\ra +|\uparrow00\ra) $. The entropy uncertainty game is only between Alice and Bob, so Charlie has been traced. When Bob gets closer to the horizon, his uncertainty about Alice's state gradually increases for both $U_1$ and $U_2^{x y}$. Given $\Omega=\omega/T_H=10$ or $30$, $U_2^{x y}$ is always better than $U_1$. For different $\Omega$ and $R_0$, $\Delta_2$ is constant while $\Delta_1$ increases as $\Omega$ and $R_0$ increase. (Colored online.)}
\label{w}
\end{figure}
We first review the definition of proper accelerated observer's vacuum states in Schwarzschild spacetime.

A Schwarzschild black hole in Schwarzschild coordinates is given by
\be
ds^2=-\bigg(1-\frac{2M}{r}\bigg)dt^2+\bigg(1-\frac{2M}{r}\bigg)^{-1}dr^2+r^2d\mathbf{\Omega}^2
\label{bh1}
\ee 
where $M$ is the mass of black hole. Near the event horizon, the metric has similar structure as Rindler horizon in flat space-time. We consider Alice and Bob free falls towards the black hole, Rob and anti-Rob hovers near the black horizon in different region. We assume Rob only detects the mode of Bob. The Penrose diagram of the Schwarzschild spacetimes is plotted in Fig.\ref{kruskal}.

We consider Dirac field for simplicity because there is at most one particle for each spin in one mode due to Pauli's exclusion principle \cite{MartinMartinez:2010ar}:  
\begin{align}\label{notation2}
 \ket{\sigma_{{\omega_i}}}_\text{I}
 &=c^\dagger_{\text{I},{\omega_i},\sigma}\ket{0}_\text{I},\nonumber\\
  \ket{\sigma_{{\omega_i}}}_\text{IV}
  &=d^\dagger_{\text{IV},{\omega_i},\sigma}\ket{0}_{\text{IV}},\nonumber\\
 \ket{\pa_{{\omega_i}}}_\text{I}
 &=c^\dagger_{\text{I},{\omega_i},\uparrow}
 c^\dagger_{\text{I},{\omega_i},\downarrow}\ket{0}_\text{I}
 =-c^\dagger_{\text{I},{\omega_i},\downarrow}c^\dagger_{\text{I},{\omega_i},
 \uparrow}\ket{0}_\text{I},\nonumber\\
\ket{\pa_{{\omega_i}}}_{\text{IV}}
&=d^\dagger_{\text{IV},{\omega_i},\uparrow}d^\dagger_{\text{IV},{\omega_i},
\downarrow}\!\ket{0}_{\text{IV}}=-d^\dagger_{\text{IV},{\omega_i},\downarrow}
d^\dagger_{\text{IV},{\omega_i},\uparrow}\!\ket{0}_{\text{IV}},
\end{align}
where $p_{{\omega_i}}$ represents a pair of spin states in the mode with
frequency ${\omega_i}$, $\sigma= \uparrow \text{or} \downarrow$, and $c^\dagger_{\text{I},{\omega_i},\sigma},d^\dagger_{\text{IV},{\omega_i},\sigma}$ are create operators for particle and anti-particle, respectively. Thus, for each mode, a Dirac particle has four basis states: $\ket{0},\ket{\uparrow},\ket{\downarrow},\ket{p}$, instead of two as given in \cite{Feng:2015xza}: $\ket{0},\ket{1}$.

The vacuum corresponding to freely falling observer is called Hartle-Hawking vacuum $|0\ra_{\text{H}}$, which is analogous to Minkowski vacuum. The vacuum corresponding to proper accelerated observer is called Boulware vacuum $|0\rangle_{\text{R}}$, which is analogous to the Rindler vacuum. There is another Boulware vacuum $|0\rangle_{\bar{\text{R}}}$ in region IV.
Vacuum is made of different frequency modes $|0_H\ra\equiv\bigotimes_i|0_{\omega_i}\ra_H$ and similarly for first excitation $|1_H\ra\equiv\bigotimes_i|1_{\omega_i}\ra_H$. The relation between different notation is
\begin{equation}\label{identif}
\begin{array}{lclcl}
\ket{0}_\text{R}&\leftrightarrow&\ket{0}_\text{I},\\
\ket{0}_{\bar{\text{R}}}&\leftrightarrow&\ket{0}_{\text{IV}},\\
\ket{0}_{\text{A,B}}&\leftrightarrow&\ket{0}_{\text{H}}.
\end{array}
\end{equation}

Just like the case in Rindler space-time, vacuum and excited state for different observer are related by Bogoliubov transformation \cite{MartinMartinez:2010ar,Alsing:2006cj,MartinMartinez:2010ds}
\begin{align}\label{vacuumf}
 \ket{0_{\omega_i}}_{\text{H}} &= (\cos q_{\text{d},i})^2
 \bikete{0_{\omega_i}}{0_{\omega_i}}\nonumber\\
 &
 +\sin q_{\text{d},i}\cos q_{\text{d},i}
 \left(\ket{\uparrow_{\omega_i}}_{{\text{R}}}
  \ket{\downarrow_{\omega_i}}_{\bar{\text{R}}}+
 \bikete{\downarrow_{\omega_i}}{\uparrow_{\omega_i}}\right)
 \nonumber\\
 &+(\sin q_{\text{d},i})^2\bikete{\pa_{\omega_i}}{\pa_{\omega_i}},
\end{align}
and for one particle state of Hartle-Hawking vacuum
\begin{align}\label{onepartf}
 \ket{\uparrow_{\omega_i}}_\text{H}&= \cos q_{\text{d},i}
 \bikete{\uparrow_{\omega_i}}{0_{\omega_i}}+
 \sin q_{\text{d},i}\bikete{\pa_{\omega_i}}{\uparrow_{\omega_i}},\nonumber\\
\ket{\downarrow_{\omega_i}}_\text{H}&=
\cos q_{\text{d},i} \bikete{\downarrow_{\omega_i}}{0_{\omega_i}}-
\sin q_{\text{d},i}\bikete{\pa_{\omega_i}}{\downarrow_{\omega_i}},
\end{align}
with
\begin{equation}\label{defr4}
\tan q_{\text{d},i}=\exp\left(-\frac{\Omega}{2}\sqrt{1-1/R_0}\;\right),
\end{equation}
where $R_0=r_0/R_H=r_0/2M$, $\Omega=\omega/T_H=8\pi\omega M$ and $\omega$ is the mode frequency measured by Bob. This approximation is only valid in vicinity of event horizon as mentioned above, i.e. $R_0-1 \ll 1$ \cite{MartinMartinez:2010ar}.

\paragraph{Incompatible measurements.}
We use 4-dimensional Pauli matrices for measurements \cite{Griffiths:2016}
\begin{align}
\sigma_x &\equiv\frac{1}{2}\begin{pmatrix}
 0 & \sqrt{3} & 0 & 0 \\
 \sqrt{3}& 0& 2& 0 \\
 0& 2& 0& \sqrt{3}\\
 0& 0& \sqrt{3}& 0
\end{pmatrix},\\
\sigma_y &\equiv \frac{1}{2}\begin{pmatrix}
 0 & -i\sqrt{3} & 0 & 0 \\
 i\sqrt{3}& 0& -2i& 0 \\
 0& 2i& 0& -i\sqrt{3}\\
 0& 0& i\sqrt{3}& 0
\end{pmatrix},\\
\sigma_z &\equiv \frac{1}{2}\begin{pmatrix}
 3 & 0 & 0 & 0 \\
 0& 1& 0& 0 \\
 0& 0& -1& 0\\
 0& 0& 0& -3
\end{pmatrix}.
\end{align}
For each pair of them, after calculating their eigenvectors, we can find the incompatible term $-\log c_1=-\log \max_{i_{1},i_{2}}\mid\langle u^{M_1}_{i_{1}}|u^{M_2}_{i_{2}}\rangle\mid^{2}$ is always $\log \frac{8}{3}$. Thus in the following discussion, without loss of generality, we choose $\sigma_x$ and $\sigma_y$ only.

\subsection{Results}

\paragraph{A Bell-like state.}
We consider a Bell-like state
\begin{equation}
    \label{entangleddir}\ket{\Psi}_{H}=\frac{1}{\sqrt{2}}\left(\ket{0}_A\ket{0}_B+\ket{\uparrow}_A\ket{\downarrow}_B\right).
\end{equation}

We depict its EUR lower bound for both $U_1$ and $U_2^{x y}$ and the difference between $H(M_{1}|B)+H(M_{2}|B)$ in Fig.\ref{00ud}.

\paragraph{W state}
Consider the case when Alice, Bob and Charlie initially shared a W state from perspective of inertial frame,
\be\label{w-state}
    W = \frac{1}{\sqrt{3}} (|00\uparrow \ra+|0\uparrow0\ra +|\uparrow00\ra)
\ee
We consider subsystem of Alice and Bob, so we must partial trace over degrees of freedom of Charlie. We depict EUR lower bound for both $U_1$ and $U_2^{x y}$ and their difference with $H(M_{1}|B)+H(M_{2}|B)$ in Fig. \ref{w}.

\paragraph{Comparison}
In all examples we calculated, $U_2$ is tighter than $U_1$. In addition, the figures shows that for a particular bound $U_1$ or $U_2$, when $\Omega = \omega / T_H$ is larger, the uncertainty bound is lower. This is evident since fixing the mode energy $\omega$, the larger $\Omega$ is, the lower Hawking temperature $T_H$ is, which results in more correlation which can reduce the uncertainty. Besides, there is no surprise that $\Delta_2^{x y}=H(M_1)+H(M_2)-(-\log c_1)$ is constant as it is only influenced by the choice of measurements $M_1, M_2$ and measured system $\rho_A$, not by the quantum memory $\rho_B$. We can see from these figures that $\Delta_1$ is not always constant but can decrease or increase when $R_0$ increase. This fact indicates that $U_1$ is not an ideal indicator of how the black hole would influence the entropic uncertainty(LHS). While for $U_2$, when the correlation decreases, the amount of increased uncertainty always equals to the amount of decreased correlation.

\section{Conclusion}\label{conclusion}
In this letter, we calculated two examples with Dirac field states in Schwarzschild spacetime, demonstrating that uncertainty relation generalized by Holevo quantity not only has a tighter lower bound, but reveals how the quantum memory would influence the entropic uncertainty as well. The second result has implications in experiments. Experiment only needs to be conducted for one Schwarzschild black hole to obtain entropic uncertainty. For all other Schwarzschild black holes, the entropic uncertainty can be exactly predicted using the new lower bound.


\begin{thebibliography}{99}
\bibitem{Heisenberg1927} W. Heisenberg, Zeitschrift für Physik 43, 172 (1927).

\bibitem{Robertson1929} H. P. Robertson,
Phys. Rev. \textbf{34}, 163 (1929).

\bibitem{Schrodinger1930} E. Schr\"{o}dinger,
Sitz. Preuss. Akad. Wiss. (Phys.-Math. Klasse) \textbf{19}, 296 (1930).

\bibitem{Abbott2016} A. A. Abbott, P.-L. Alzieu, M. J. W. Hall and C. Branciard,
Mathematics \textbf{4}, 8 (2016).

\bibitem{Schwonnek2017} R. Schwonnek, L. Dammeier and R. F. Werner,
arXiv:1705.10679.

\bibitem{Xiao2017I} Y. Xiao, C. Guo, F. Meng, N. Jing and M.-H. Yung,
arXiv:1706.05650.

\bibitem{Bialynicki1975} I. Bia{\l}ynicki-Birula and J. Mycielski, 
Commun. Math. Phys. \textbf{44}, 129 (1975).

\bibitem{Deutsch1983}  D. Deutsch, Physical Review Letters 50, 631 (1983).

\bibitem{Kraus1987} K. Kraus,
Phys. Rev. D. \textbf{35}, 3070 (1987).

\bibitem{Maassen1988}  H. Maassen and J. B. M. Uffink, Physical Review Letters 60, 1103 (1988).

\bibitem{Berta2010} M. Berta, M. Christandl, R. Colbeck, J. M. Renes, and R. Renner, Nature Physics 6, 659 (2010).

\bibitem{XiaoSci2016} Y. Xiao, N. Jing, and X. Li-Jost, Scientific Reports 6, (2016).

\bibitem{Feng:2015xza} J. Feng, Y.-Z. Zhang, M. D. Gould, and H. Fan, Physics Letters B 743, 198 (2015).

\bibitem{Alice:2005}
I. Fuentes-Schuller and R. B. Mann, Physical Review Letters 95, (2005).

\bibitem{Prevedel:2011}
R. Prevedel, D. R. Hamel, R. Colbeck, K. Fisher, and K. J. Resch, Nature Physics 7, 757 (2011).

\bibitem{Li:2011}
C.-F. Li, J.-S. Xu, X.-Y. Xu, K. Li, and G.-C. Guo, Nature Physics 7, 752 (2011).

\bibitem{Reza1961} C. L. Mallows and F. M. Reza, The American Mathematical Monthly 71, 108 (1964).

\bibitem{Hall1994} M. J. W. Hall,
Phys. Rev. Lett. \textbf{74}, 3307 (1994).

\bibitem{Bennett1982} C. H. Bennett,
Int. J. Theor. Phys. \textbf{21}, 905 (1982).

\bibitem{Ollivier2001} H. Ollivier and W. H. Zurek,
Phys. Rev. Lett. \textbf{88}, 017901 (2001).

\bibitem{Nielsen2011} M. A. Nielsen and I. L. Chuang, Quantum computation and quantum information (Cambridge University Press, Cambridge, 2015).

\bibitem{Marco} P. J. Coles, M. Berta, M. Tomamichel, S. Wehner, Reviews of Modern Physics, 2017.

\bibitem{Coles} P. J. Coles and M. Piani, Physical Review A 89, (2014).

\bibitem{XJ} Y. Xiao, N. Jing, S.-M. Fei, and X. Li-Jost, Journal of Physics A: Mathematical and Theoretical 49, (2016).

\bibitem{Kaniewski} J. C. Kaniewski, M. Tomamichel, and S. Wehner, Physical Review A 90, (2014).

\bibitem{Rudnicki} Ł. Rudnicki, Z. Puchała, and K. Życzkowski, Physical Review A 89, (2014).
  
\bibitem{Almheiri:2012rt} A. Almheiri, D. Marolf, J. Polchinski, and J. Sully, Journal of High Energy Physics 2013, (2013).
  
\bibitem{MartinMartinez:2010ar} E. Martín-Martínez, L. J. Garay, and J. León, Physical Review D 82, (2010).
  
\bibitem{Griffiths:2016} D. J. Griffiths, Introduction to quantum mechanics (Cambridge University Press, Cambridge, 2016).
  
  
\bibitem{Alsing:2006cj} P. M. Alsing, I. Fuentes-Schuller, R. B. Mann, and T. E. Tessier, Physical Review A 74, (2006).

\bibitem{MartinMartinez:2010ds} E. Martín-Martínez and J. León, Physical Review A 81, (2010).
  
\bibitem{Page:1993} D. N. Page, Physical Review Letters 71, 3743 (1993).
  
\bibitem{Pan:2008} Q. Pan and J. Jing, Physical Review D 78, (2008).
  
\bibitem{Coffman:2000} V. Coffman, J. Kundu, and W. K. Wootters, Physical Review A 61, (2000).
  
\bibitem{Osborne:2006} T. J. Osborne and F. Verstraete, Physical Review Letters 96, (2006).
  
\bibitem{Daniel:2016} D. Harlow, Reviews of Modern Physics 88, (2016).
  
\end{thebibliography}
\end{document}